\documentclass[aps,pre,10pt,twocolumn,floatfix,nofootinbib,superscriptaddress]{revtex4-1}

\usepackage{amssymb}
\usepackage{amsmath}
\usepackage{amsthm}
\usepackage{graphicx}

\usepackage{xcolor}
\usepackage[utf8]{inputenc}
\usepackage[unicode=true]{hyperref}
\usepackage[para,online,flushleft]{threeparttable}
\hypersetup{
    colorlinks=true,
    linkcolor=blue,
    urlcolor=blue,
    citecolor=blue
}
\renewcommand{\vec}{\mathbf}
\begin{document}
\title{The Multiscale Principle in phase transition: Geometric structure, thermodynamic entropy, and dynamical instability }

\author{Patricio Venegas-Aravena}
\email{plvenegas@uc.cl}
\affiliation{Applied Physics Group, Department of Engineering, \emph{Universidad Loyola Andaluc\'ia}, Avda. de las Universidades s/n, 41704, Dos Hermanas, Seville, Spain}
\author{Fernando A. Oliveira}
\email{faooliveira@gmail.com}
\affiliation{International Center of Physics, Institute of Physics, \emph{University of Brasilia}, 70910-900, Brasilia, Federal District, Brazil}

\begin{abstract}
Scaling and multi-scaling are two guiding principles for understanding nature or even commerce and industrial outcome. In the realm of critical phenomena, scaling laws describe the behavior of physical quantities near a continuous phase transition. This is a complex phenomena where systems exhibit critical fluctuations and long-range correlations, challenging classical formulations such as the fluctuation-dissipation theorem (FDT). In particular, near critical points, the divergence of the correlation length and the emergence of fractal geometry in spin clusters suggest the need for alternative frameworks. In this work, we explore the connection between critical exponents, fractal dimensions, and the thermodynamic formalism known as the Principium Luxuriae. By linking the thermodynamic fractal dimension $D$ to the correlation length and entropy variations, we reinterpret energy dissipation in terms of multiscale processes. This approach allows us to reproduce effective scaling tendencies associated with the critical behavior of the 2d Ising model and provides new insight into how systems redistribute energy macroscopically during phase transitions.  Furthermore, we demonstrate that the decrease in $D$ near the critical temperature marks a transition toward macroscopic coherence, where the system begins to dissipate energy across its entire scale. This unified behavior, however, induces a heightened sensitivity to perturbations, where local fluctuations are transmitted globally through the fractal network. This mechanism establishes a direct bridge between geometric structure, thermodynamic entropy, and the onset of dynamical instability (chaos) as the system approaches criticality.

\end{abstract}
\maketitle
\section{Introduction}
\label{Int}

Scaling principles have long served as a cornerstone in our comprehension of natural phenomena, extending their utility across disciplines from the hard sciences to the complexities of socio-economic systems \cite{Stanley99, West17}. Within the realm of physics, the study of critical phenomena, particularly continuous phase transitions, has been profoundly shaped by the identification and analysis of scaling laws \cite{Cardy96,Goldenfeld18}. Near critical points, systems exhibit a rich tapestry of behaviors characterized by diverging correlation lengths, power-law relationships between physical quantities, and the emergence of fractal structures in correlated domains (e.g. \cite{Cabrera21}). These features often challenge the applicability of classical statistical mechanics frameworks, such as the fluctuation-dissipation theorem (FDT), which typically relates the response of a system to an external perturbation with its equilibrium fluctuations \cite{WuWang20}.

The Ising model, in its various dimensionalities, has served as a quintessential system for investigating critical phenomena and the associated scaling behaviors (e.g. \cite{fontana19}). Its apparent simplicity belies a profound complexity near the critical temperature, where long-range correlations and scale invariance manifest. Traditional theoretical approaches based on Euclidean geometry often encounter limitations in fully capturing the observed critical exponents and the complex spatial organization of fluctuations \cite{Fisher64}. For instance, the mean-field treatment of the Ising model yields predictions that deviate significantly from experimental and high-precision numerical results, particularly concerning the critical exponent $\eta$, which governs the anomalous dimension of the order parameter correlation function at criticality \cite{Cardy96,Fisher64,Lima24,Lima25}.

The breakdown of the FDT in various complex systems, including but not limited to critical phenomena, structural glasses \cite{Grigera99,Ricci-Tersenghi00}, disordered magnetic systems \cite{Vainstein05}, and even non-equilibrium systems governed by the Kardar-Parisi-Zhang (KPZ) universality class \cite{Kardar86}, has spurred the exploration of alternative theoretical frameworks. A recurring theme in these investigations is the potential role of non-Euclidean geometries, particularly fractal geometry, in describing the underlying structure and dynamics of these systems (e.g., \cite{Katzav03}). The concept of a fractal dimension, characterizing the space-filling capacity of a set, offers a powerful tool to quantify the self-similar structures that emerge near criticality and in other complex scenarios.

Fractional calculus, which extends the notion of differentiation and integration to non-integer orders, provides a mathematical framework to naturally incorporate fractal concepts into physical descriptions \cite{Macias-Diaz22}. The Riesz fractional derivative, in particular, has been employed to model phenomena occurring on fractal substrates or exhibiting anomalous scaling behaviors (e.g., \cite{chen10}). Its connection to fractal dimensions and its ability to capture non-local interactions make it a promising tool for addressing the limitations of Euclidean-based approaches in critical phenomena.

In parallel to these developments in statistical physics, the field of thermodynamics has also seen extensions aimed at understanding energy dissipation and organization in complex systems across multiple scales. The Principium Luxuriae, a tenet of multiscale thermodynamics, posits that systems driven away from equilibrium tend to dissipate excess energy across a spectrum of scales, guided by a thermodynamic fractal dimension that reflects the balance between large-scale coherence and small-scale fluctuations \cite{Venegas-Aravena2022a,Venegas-Aravena2024}. This framework offers a geometric interpretation of energy dissipation processes, linking the structural properties of a system to its thermodynamic behavior.

Building upon these established lines of research, this work aims to bridge the gap between the geometric characterization of critical phenomena, particularly through the lens of fractal dimensions derived from correlation functions, and the thermodynamic description offered by the Principium Luxuriae. By employing the framework of fractional differential geometry, we explore how the emergence of fractal structures near the critical point of the 2d Ising model can be related to the multiscale dissipation of energy as quantified by the thermodynamic fractal dimension. Specifically, we investigate the connection between critical exponents, such as $\eta$ and $\nu$, and the thermodynamic fractal dimension $D$, seeking to provide a novel interpretation of energy redistribution during phase transitions. Furthermore, we explore the implications of a changing fractal dimension near the critical temperature for the system's dynamical behavior, particularly in relation to the onset of chaos and large-scale coherence.

The subsequent sections of this paper are organized as follows: Section \ref{GC} delves into general concepts, focusing on the correlation function and the challenges posed by the critical point to traditional Euclidean formulations. Section \ref{MSTh} introduces the framework of multiscale thermodynamics and the Principium Luxuriae, outlining the role of the thermodynamic fractal dimension in characterizing energy dissipation, and also does some applications to the $2d$ Ising model. Section \ref{disc} presents a discussion of the connections established between the fractal dimension derived from the Ising model and the thermodynamic fractal dimension, along with their implications for the system's behavior near criticality. Finally, Section \ref{concl} summarizes the key findings and outlines potential avenues for future research.

\section{General Concepts}
\label{GC}

 Our investigation commences with the correlation function $G(r)=\langle \psi(\vec{r}+\vec{i})\psi(\vec{i})\rangle$, where $\psi (\vec{i})$ represents the fluctuation of the order parameter at position $\vec{i}$. This crucial function $G(r)$  is known to obey the equation ~\cite{Lima24}
\begin{equation}
(-\nabla^2 +\rho^{-2})G(r)=\delta^{d_E}(r),
\label{G}
\end{equation}
Here, $d_E$ denotes the Euclidean dimension of the embedding space, and $\rho$  signifies the correlation length, a measure of the spatial extent over which fluctuations are correlated ~\cite{Cardy96}. The mean-field solution to Eq. (\ref{G})  is given by
\begin{equation}
\label{G01}
    G(r)=r^{2-d_E} \exp(-r/\rho).
\end{equation}
However, this solution falters at the critical point, where the correlation length $\rho$ diverges following a power law, $\rho \propto |T-T_c|^{-\nu}$.
To reconcile this theoretical prediction with experimental observations, an empirical correction involving the critical exponent $\eta$ is typically introduced ~\cite{Fisher64},

\begin{equation}
\label{G2}
G(r) \propto  r^{2-d_E-\eta}.
\end{equation}
Correlation functions, as highlighted in the review by Gomes Filho et al. ~\cite{GomesFilho25}, embody a fundamental form of the fluctuation-dissipation relation (FDR), readily linked to the susceptibility ~\cite{Goldenfeld18}.
Intriguingly, the FDR encounters limitations at critical points, failing to inherently explain the form of Eq. (\ref{G2}) in the short-distance regime $ r \ll\rho $. The emergence of the Fisher exponent $\eta$ reveals the need to ensure consistency with the scaling relations between critical exponents, a prime example being the Fisher scaling relation ~\cite{Cardy96}.
\begin{equation}
\label{gamma0}
\gamma =(2-\eta)\nu,
\end{equation}
where $\gamma$ is the susceptibility exponent and $\nu$ is the critical exponent governing the divergence of the correlation length. Indeed, the response function Eq. (\ref{G}) within the standard Euclidean formulation predicts $\eta=0$ a clear contradiction to the experimental findings, thus illustrating a breakdown of the FDR.
As noted earlier, correlation functions represent a simplified manifestation of the fluctuation-dissipation relation ~\cite{GomesFilho25},intrinsically connected to susceptibility~\cite{Goldenfeld18}.
The violation of the FDR, as exemplified by Eq. (\ref{G2}), has been attributed to a fundamental shift in the underlying geometry, transitioning from a Euclidean to a fractal structure ~\cite{Lima24}. This breakdown of FDR is a well-documented phenomenon in various systems, including structural glasses ~\cite{Grigera99,Ricci-Tersenghi00,Crisanti03,Barrat98,Bellon02,Bellon06}, the random exchange Heisenberg chain ~\cite{Vainstein05}, proteins~\cite{Hayashi07},  mesoscopic radioactive heat transfer ~\cite{Perez-Madrid09,Averin10},  ballistic diffusion \cite{Costa03,Costa06,Lapas07,Lapas08}, and the Kardar-Parisi-Zhang (KPZ) universality class ~\cite{Kardar86,Rodriguez19,Anjos21,GomesFilho21,GomesFilho24}. It is quite important as well to call attention that in growth phenomena we have well-defined scaling properties~\cite{Family85,Rodrigues24,Rodriguez25,Moca25}.
 A common thread underlying the failure of the FDR is the breaking of ergodicity \cite{GomesFilho21,Costa03,Wen23}.

Consequently, as the system approaches the critical temperature ($T \rightarrow T_c$), thermodynamic functions become power laws, as do response functions~\cite{Cardy96,GomesFilho24},  the correlation length diverges ($\rho  \rightarrow  \infty$), thus exhibiting scale invariance, which turns Eq.(\ref{G}) into the power law Eq. (\ref{G2}), closing the cycle and showing that all these behaviors are connected. To address the limitations of the Euclidean formulation, we employ the framework of fractional differential geometry \cite{Muslih10,Muslih10b}, specifically utilizing the Riesz fractional derivative of order $\zeta$, where $1/2<\zeta<1$. This derivative is associated with the Riesz fractal dimension $d_R$, satisfying $d-1<d_R<d$.  In this context, Eq.(\ref{G}) is replaced by:
\begin{equation}
\label{SG3}
(-\nabla^2)^\zeta \left(\frac{1}{|\vec{r}-\vec{r'}|^{d_R-2\zeta}} \right) = \frac{4^{\zeta} \pi^{d_R/2} \Gamma(\zeta)}{ \Gamma(d_R/2-\zeta)} \delta^{d_R}(\vec{r}-\vec{r'}),
\end{equation}
which leads to the modified equation for the correlation function:
\begin{equation}
\label{G3}
(-\nabla^2)^\zeta G(r)=\delta^{d_R}(r),
\end{equation}

This modification reflects the constraint of dynamics in a fractal structure characterized by the correlation fractal dimension $d_R$. For simplicity, we have set the proportionality constant to unity. The solution of Eq. (\ref{G3}) naturally recovers Eq. (\ref{G2})) with a nonzero exponent ~\cite{Lima24}:
\begin{equation}
\label{eta0}
 \eta=d_E-d_R,
\end{equation}
where $d_R$ is interpreted as the Riesz fractal dimension, arising from the enhanced correlations within clusters at the phase transition, or the fractal correlation dimension itself
 ~\cite{Lima24,Lima25,Carrasco26}. Notably, the relations (\ref{eta0}) and (\ref{G2}) emerge organically within this framework, without the need for ad hoc assumptions. This represents an exact solution for $\eta$ ~\cite{Lima24}. In this manner, the fractional formulation provides an effective geometrical framework capable of reproducing the observed scaling behavior beyond the standard Euclidean FDT description. Subsequently~\cite{Lima25,Carrasco26}, these relationships were verified for non-integer dimensions in the range $1 \leq d \leq 4$. Furthermore, this has also been verified for disordered Ising systems~\cite{lima26}.

Next, we successfully examine the predictions obtained for several fundamental thermodynamic models: for the Ising and Potts models, the approach works perfectly in all dimensions. It is also possible to describe the XY and Heisenberg systems~\cite{Souza26} for $d \neq 2$. For $d=2$, the XY and Heisenberg models do not exhibit a thermodynamic phase transition; therefore, our approach does not apply to this case. \\

Having established a framework rooted in fractional differential geometry to address the scaling behavior and the limitations of the classical Euclidean fluctuation-response formulation near critical points, we now turn our attention to the thermodynamic implications of the fractal correlation structure. In particular, although the relation between the anomalous exponent $\eta$ and the fractal dimension $d_R$ emerges naturally within the fractional framework, an additional challenge remains: establishing how such geometric scaling connects to the singular thermodynamic behavior characteristic of continuous phase transitions. Since the divergence of the specific heat and free-energy derivatives near $T_c$ originates from singular entropy contributions, the following section explores whether the thermodynamic fractal dimension $D$ can provide an effective multiscale description linking correlation geometry, entropy transfer across scales, and critical thermodynamic behavior. From a thermodynamic perspective, the emergence of a fractal correlation structure implies that entropy exchange is no longer distributed uniformly over an Euclidean volume, but instead propagates through a scale-dependent correlated geometry. Consequently, the divergence of the correlation length near criticality suggests that thermodynamic transfer itself acquires a multiscale character, motivating the introduction of an effective thermodynamic fractal dimension associated with entropy redistribution across correlated domains.

\section{Multiscale thermodynamics: Application to the two-dimensional Ising model}
\label{MSTh}
From Equations (\ref{gamma0}) and (\ref{eta0}), an approximate inverse tendency is observed between the critical exponent and the fractal dimension, as follows:

\begin{equation}
\label{eq:5}
 d_R \propto \nu^{-1}
\end{equation}
{In Table \ref{Table1} we show some values of $d_R$ and $\nu$ for the Ising model. Note that for higher dimensions the above  approximation becomes better. } The critical exponent governs the manner in which systems approach the critical point that determines a phase transition. During this approach, the scaling of various physical properties is obtained. In other words, a critical exponent establishes power-law relationships between various parameters in the vicinity of the critical point.

The Principium Luxuriae, a branch of multiscale thermodynamics, posits that systems, in response to external forces, tend to release excess energy across multiple scales when seeking a new equilibrium state \cite{Venegas-Aravena2022a, Venegas-Aravena2024}. The parameter defining this principle is the thermodynamic fractal dimension ($D$), which can be interpreted as a balance between large-scale and small-scale energy dissipation. In other words, the geometric properties of the system are intrinsically related to the multiscale dissipation of energy. Mathematically, this parameter is expressed as \cite{Venegas-Aravena2022b,Venegas-Aravena2024}:

\begin{equation}
\label{eq:ms1}
 D = - k_{V} \ln{\omega_0 \frac{dS}{dS_0}}
\end{equation}

Where $k_{V} =1/(\ln\frac{r}{r_0})$, $r$ represents the system size, $r_0$ denotes the reference size (associated with the size of the smallest subdomain of the system), $\omega_0 = e^{(1-d_E)/k_V}$ and where $dS$ and $dS_0$ signify the macroscopic and microscopic change of entropy, respectively.

Low $D$ values are associated with the emergence of spatially coherent structures, indicating that large regions of a system exhibit similar energy states. This suggests that the system, in seeking to release energy macroscopically, adopts geometric configurations characterized by a lower fractal dimension. Conversely, high $D$ values are associated with more spatially complex configurations.

The Ising model indicates that, at low temperatures, the system tends towards order, with spins aligning in the same direction, resulting in macroscopic magnetization. As temperature increases, this order is lost, and the system becomes paramagnetic. The phase transition between these two states occurs at the critical temperature ($T_c$). Near this point, the system exhibits critical fluctuations, where regions of aligned spins coexist with regions of randomly oriented spins. It is precisely within this critical regime that fractal dimension plays a crucial role. The fractal dimension of aligned spin clusters (or magnetic domains) near $T_c$ has been shown to be less than the dimension of the embedding space (which is 2 in this case).

The connection with the Principium Luxuriae arises from considering that, near the phase transition, the system experiences significant fluctuations in spin orientation. These fluctuations can be interpreted as a response to "external forces" (the application of heat to the system) that perturb the system's equilibrium. Here, the low $D$ values in the Principium Luxuriae correspond to the presence of large domains of aligned spins in the $2d_E$ Ising model. That is, low $D$ values would be achieved at lower temperatures. However, a low $D$ in the Principium Luxuriae is associated with lower temperatures when there is no macroscopic dissipation since $dS_0 \propto T$ for a constant $dS$. This suggests that $dS$ must not be constant for the Ising model case. Considering Equation \ref{eq:ms1}, this variation is described by:

\begin{equation}
\label{eq:ms2}
 dS = T \frac{C_0}{\omega_0} e^{-D/k_V}
\end{equation}
Where $C_0$ is a constant with dimensions of energy (temperature). Within the framework of the Principium Luxuriae, it is posited that fractal dimension serves as an indicator of which scale (large or small) dominates energy dissipation. Specifically, in the case of a system approaching the critical temperature, the effective thermodynamic fractal dimension (in this case $D \approx d_R$ is assumed) evolves toward lower values near criticality  (Ref. \cite{Lima24}). Thus, motivated by the approximate scaling tendency observed between $d_R$ and $\nu$, we phenomenologically assume an effective correspondence between the thermodynamic fractal dimension and the correlation-length scaling exponent, namely $D^{-1}\sim \nu$, the correlation length may be phenomenologically rewritten as:
\begin{equation}
\label{eq:cl}
 \rho \sim |T-T_c|^{-1/D}.
\end{equation}

Importantly, Equation~\ref{eq:cl} should not be interpreted as a fundamental replacement of the exact Ising exponent $\nu$, but rather as an effective phenomenological parametrization linking the thermodynamic fractal dimension to the dominant scaling length near criticality. In this sense, the relation $1/D \sim \nu$ is introduced only as an asymptotic multiscale tendency emerging from the entropy-transfer structure, and not as a strict renormalization-group identity.

\begin{figure}[htbp]
	\centering
	\includegraphics[width=1\columnwidth]{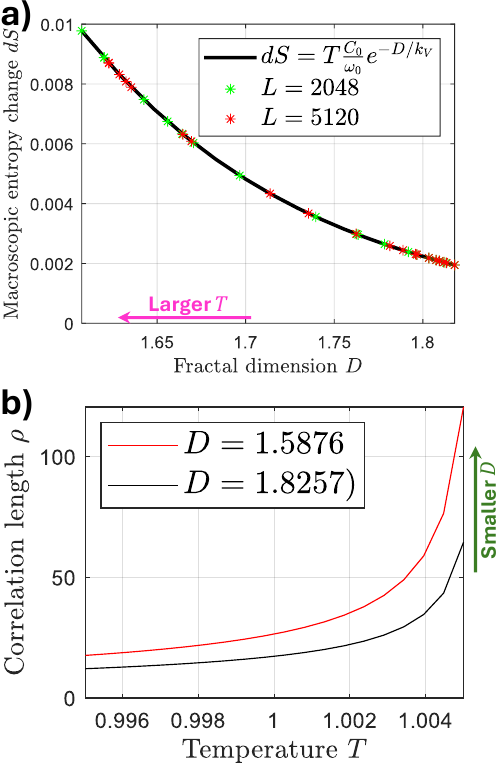}
        \caption{a) Relationship between temperature, fractal dimension, and macroscopic entropy for the Ising model. The black curve illustrates the function described by Equation \ref{eq:ms2} for the values $r = 2048$, $r_0 = 1$, $d_E = 2$, and $C_0 = 1$. The green and red curves correspond to data from Refs. \cite{Lima24,Lima25}. b) increasing correlation length by decreasing $D$ by using Equation \ref{eq:cl}.}
	\label{fig:1_ms}
\end{figure}

This suggests that spin disordering may be a less prominent effect, as the system must be undergoing more dominant macroscopic changes than spin disordering alone. This can be observed in Figure \ref{fig:1_ms}a, where an increase in temperature is associated with an increase in $dS$ when using Equation \ref{eq:ms2}. An explanation for the increase in large-scale dissipation can be related to the increase in the correlation length $\rho$ of the spins as shown in Figure \ref{fig:1_ms}b (using Equation \ref{eq:cl}). In this case of greater correlation length, the fluctuations of the system encompass a greater number of spins and therefore extend to greater distances. This effect can generate changes in large portions of the system, potentially having effects at macroscopic scales.

\begin{table*}[t]
\begin{threeparttable}

\begin{tabular}{c|ccc} \hline\hline
$d_E$                            & $\nu$                                                   & $d_R$ & $\nu d_R$  \\ \hline
4                           & 1/2                            &  4 & 2   \\
3.75  & $0.523405 (15)$  &  $3.74856(4)$ &  $1.9620(6)$  \\
3.5    & $0.55215(2)$    &  $3.493163(3)$ &  $1.9287(1)$ \\
3.25     & $0.5873(10)$    &  $3.2321(20)$ & $1.898(3)$   \\
3       & $0.6310(15)$    & $2.9635(25)$ & $1.870(6)$   \\
2.75    & $0.686(3)$      &  $2.6836(9)$ & $1.841(9)$   \\
2.5      & $0.758(5)$      & $2.3918(8)$ & $1.813(13)$   \\
2.25   & $0.857(6)$      & $2.082(1)$ & $1.784(13)$    \\
2        & 1       & $7/4$ &  7/4     \\ \hline\hline
\end{tabular}
\caption{ Values of critical exponents for the Ising model in   $d_E$ dimensional lattice. For integer values of $d_E$ we get $\nu$ from refs \cite{Coniglio89,Pelissetto02} for fractional values from \cite{Guillou87}, $d_R$ from \cite{Lima24,Lima25}, in the last column we check out the approximation $ \nu \approx d_R^{-1}$,
Eq.  (\ref{eq:5}).}
\label{Table1}

\end{threeparttable}
\end{table*}

\begin{figure}[htbp]
	\centering
	\includegraphics[width=1\columnwidth]{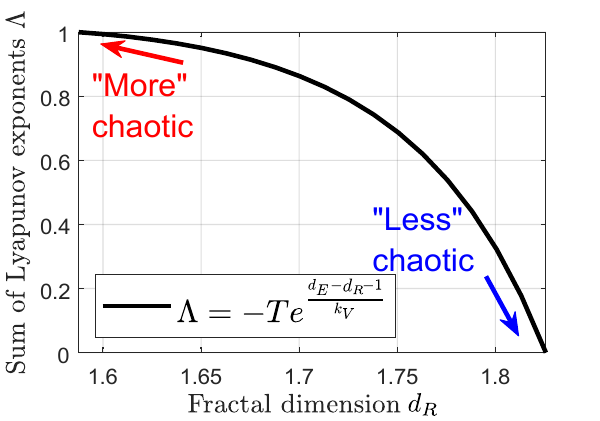}
        \caption{Sum of the Lyapunov exponent in terms of $d_R(T)$ by using Equations from Ref. \cite{Lima24,Venegas-Aravena2024}. The figure illustrates an effective increase in dynamical sensitivity and fluctuation amplification as the system approaches the critical temperature within the multiscale thermodynamic framework.}
	\label{fig:2_ms}
\end{figure}

\section{Discussions}
\label{disc}

The phase transition observed in the 2d Ising model provides an ideal context for exploring the similarities between $D$ and the fractal dimension $d_R$ observed in previous models in the vicinity of the critical temperature $T_c$ \cite{Lima24}. The convergence of $D$ and $d_R$ toward decreasing values as the temperature approaches $T_c$ suggests a macroscopic manifestation of energy release (Figure \ref{fig:1_ms}a), within the framework of multiscale thermodynamics, driven by the increase in the spin correlation length. This phenomenon implies that spins can exert mutual influence at considerable (or macroscopic) distances, which is reflected in the inverse relationship between $D$ and the correlation length (Figure \ref{fig:1_ms}b). In this sense, an increase in correlation length could be viewed as a macroscopic manifestation of the release of external energy, which is driven by the temperature increase. This large-scale dissipation process prepares the system for the phase transition, where the microscopic order gives way to a different macroscopic state.

This interpretation in the multiscale thermodynamic context is supported by the chaotic behavior observed in the system near the critical temperature \cite{Baity-Jesi21,Ney-Nifle97}. The Equation 6 in ref. \cite{Venegas-Aravena2024} establishes a relationship between $D$ and the sum of the Lyapunov exponent $\Lambda$, arguing that the greater the $\Lambda$, the larger the chaotic features of the systems. After considering $dS_0 \propto T$ and $d_R(T)=7/4+a(T-T_c)+b(T-T_c)^2$, this Equation becomes:

\begin{equation}
\label{eq:ms3}
 \Lambda = -T e^{(d_E-d_R(T)-1)/k_V}
\end{equation}
The behavior of this Equation can be seen in Figure \ref{fig:2_ms}, which demonstrates that a decrease in $d_R$ is correlated with an increase in $\Lambda$, suggesting that the system becomes more chaotic as it approaches $T_c$. This increase in chaos reflects the growing complexity and fluctuations of the system in the phase transition. The agreement between a black curve, derived from a multiscale thermodynamic framework, and empirical data (Figure \ref{fig:1_ms}a) suggests the validity of multiscale physics within the 2d Ising model. In this instance, both the qualitative and quantitative behavior of $D$ near $T_c$ exhibits similarities to that of $d_R$. This observed correspondence provides a basis for establishing a connection between $D$ and $d_R$.

However, a subtle point regarding the interpretation of $D$ must be addressed to avoid a potential misunderstanding. In previous macroscopic applications of the Principium Luxuriae (such as in the study of seismic activity by Venegas-Aravena and Zaccagnino \cite{Venegas-Aravena2025}) a higher thermodynamic fractal dimension $D$ is often associated with higher complexity and chaotic dynamics in systems that are inherently out of equilibrium. In contrast, our results for the Ising model show that a lower $D$ is what signals the progression toward chaos and the amplification of perturbations near the critical point. This is not a contradiction, but rather a reflection of the different roles energy dissipation plays in these systems. In the case of the 2d Ising model, a lower $D$ indicates that the system is successfully shifting its energy dissipation from microscopic to macroscopic scales. Near the critical temperature, this macroscopic coherence means that the system is no longer a collection of independent parts; instead, it acts as a unified entity where the correlation length is so vast that any local perturbation is transmitted globally. Therefore, the ``chaos'' observed here refers to this systemic hypersensitivity, a direct consequence of the system's attempt to redistribute energy at the largest possible scale as it prepares for the phase transition. In this sense, while in other systems a high $D$ reflects a fragmented, complex dissipation, in the critical Ising model, a low $D$ represents a ``causal'' or ``coherent'' chaos, where the breakdown of the fluctuation-dissipation theorem is explained by the emergence of a macroscopic response to microscopic changes.

The additional fact that using $d_R$ and relating it to chaotic behaviors that have been investigated for this model can ensure an approximation between $D$ and $d_R$. This motivates the phenomenological correspondence:
\begin{equation}
\label{eq:ms4}
 D \sim d_R
\end{equation}
which should be interpreted only as an effective asymptotic tendency emerging near criticality, rather than as an exact equality between the thermodynamic and geometric fractal dimensions. However, during a phase transition, the entanglement entropy of a block of spins exhibits a characteristic behavior: it increases with the size of the block until it saturates, and this saturation is determined by the dynamic correlation length of the system (e.g. \cite{Cincio2007}). This suggests a strong connection between the concept of entropy, the size of the system's subdomains, and the correlation length. In this context, an important point to highlight is the interpretation of entropy in such a system. In the case of the Principium luxuriae, a smaller thermodynamic fractal dimension indicates that the system predominantly releases energy at larger scales, meaning that more extensive portions of the system are involved in the energy release. This implies a scaling relationship between entropy change and the scale at which it occurs, which is consistent with the increase in correlation length observed for lower values of the fractal dimension. Since a longer correlation length is associated with enhanced collective fluctuations and larger correlated domains, this description provides a natural link between the fractal dimension of the Ising model ($d_R$) and the thermodynamic fractal dimension ($D$) of multiscale thermodynamics. This unified approach is particularly relevant because it circumvents the traditional need to decompose entropy into a normal part ($S_n$) and a singular part ($S_{sing}$). Nevertheless, to establish consistency with the standard theory of critical phenomena, Appendix A explicitly reformulates the multiscale entropy contribution in terms of the singular entropy sector $S_{sing}$ and examines how the associated logarithmic specific-heat divergence may emerge effectively from the scaling structure of the thermodynamic fractal dimension. In our framework, the singular behavior observed at the phase transition is not an added term, but an emergent property of the scaling parameter $r$ when it scales with the correlation length $\rho$ near $T_c$.  Despite these correspondences, additional work is still required to establish a rigorous derivation of the singular thermodynamic contributions directly from the multiscale framework. In particular, while the present approach captures effective scaling relations associated with criticality, a complete derivation of the non-analytic entropy contribution remains an open problem. Nevertheless, the observed consistency between the scaling structure of the Ising model and the thermodynamic fractal dimension suggests that multiscale thermodynamics may provide an effective geometrical framework for interpreting critical phenomena. At present, the framework should therefore be interpreted as a phenomenological bridge between fractal correlation geometry and thermodynamic critical scaling, rather than as a first-principles derivation of the singular free-energy structure.

\subsection{Insights from the Specific Heat}
The framework developed in this work, particularly Equation~\ref{eq:ms2}, can be extended to describe second-order phase transitions by identifying the thermodynamic fractal dimension $D$ with a system-specific exponent and interpreting the scale parameter $r$ as the size of the region influenced by the correlation length. This reinterpretation connects the geometric entropy structure to standard critical phenomena, in which diverging correlation lengths produce singular behavior in thermodynamic observables.
Particularly, Appendix A presents an analytical justification of this correspondence by starting from the classical divergence of the specific heat near the critical temperature, $C_H \propto \ln|T - T_c|$.
In the $2d$ Ising universality class, this logarithmic divergence arises from marginality at the upper critical dimension of the entropy fluctuations, and therefore cannot be represented exactly by a power-law scaling form. The use of an effective exponent $\xi$ should thus be understood as a local reparametrization of the logarithmic singularity rather than a true asymptotic equivalence. Under this constrain, the behavior is shown to be approximated by a power law of the form $|T - T_c|^\xi$ for small $\xi$. This leads to a generalized expression for the entropy derivative, $\partial S / \partial T \sim |T - T_c|^\xi$, which can be compared to the result derived from Equation~\ref{eq:ms2} under the assumption that $r(T) \sim |T - T_c|^\mu$.
The comparison yields a direct relation between the scaling exponent $\xi$, the correlation length exponent $\nu$, and the thermodynamic fractal dimension $D$: $
\xi = \nu D - 1$.
This shows that $\xi$ encapsulates how entropy scaling near criticality is constrained by both thermodynamic geometry and correlation growth. In this view, Equation~\ref{eq:ms2} serves as a generalization of the critical behavior, where the appearance of singularities corresponds to specific values of $\xi$ within a broader, continuous entropy framework. As such, the formal structure remains consistent while allowing for variations that recover known critical exponents in limiting cases.
Moreover, the differences across the systems listed in Table~\ref{Table1}, which show variations in effective exponents and entropy gradients, can be attributed to small but nonzero changes in $\xi$. Since $\xi$ is derived from scaling behavior and dimensional coupling, it need not be fixed; it reflects the local multiscale thermodynamics of each system. For instance, if we set $D\approx d_R$ (particular case when $\xi \sim  0$), we can investigate how $\xi$ adjusts. For $d_E=2$, with $\nu=1$ and $d_R=7/4$, implies that $\xi=0.75$. This value of $\xi$ is clearly distinct from the formal logarithmic limit $\xi\rightarrow 0$. Instead, it reflects an effective scaling parametrization within the multiscale framework, showing how the entropy-transfer structure adjusts phenomenologically to accommodate the observed Ising exponents. This indicates that the assumption $D\approx d_R$ alongside the derived relation for $\xi$ provides an effective scaling behavior, even if it doesn't precisely yield $\xi=0$ for the 2D Ising model's exact logarithmic divergence. The framework thus provides a consistent scaling relationship that, while not perfectly matching the $\xi=0$ case in a direct power-law sense, offers a general description where $\xi$ can vary to fit different critical behaviors. Therefore, the Principium Luxuriae (as expressed through the general Equation~\ref{eq:ms2}) provides a unified description in which phase transitions correspond to the limits where entropy transfer across scales is maximized or diverges, governed by the structure of $\xi$, $\nu$, and $D$.

\section{Conclusions}
\label{concl}

The observed convergence of the thermodynamic dimension $D$ and the fractal dimension $d_R$ near the critical temperature $T_c$ in the 2d Ising model supports a multiscale thermodynamic interpretation of the phase transition. That is, both quantities exhibit similar effective scaling tendencies near criticality. The inverse relationship between $D$ and the spin correlation length (given by Equation \ref{eq:cl}) indicates that increasing long-range interactions, driven by rising temperature, facilitate macroscopic energy dissipation, preparing the system for the transition. Furthermore, the correlation between the decrease $d_R$ and the increase in the Lyapunov exponent $\Lambda$, as described by Equation \eqref{eq:ms3},suggests that the system exhibits heightened chaotic behavior as it approaches $T_c$ reflecting the increasing complexity and fluctuations. The agreement between the multiscale thermodynamic framework and empirical data for $D$ near $T_c$, along with the consistency between the behaviors of $D$ and $d_R$, suggests a phenomenological connection between these two dimensions.   Building upon these insights, we aim to extend these results to the study of networks in non-integer dimensions~\cite{Lima25}, disordered networks~\cite{Vermeulen17}, and dynamic phase transitions~\cite{Santos24,Gutierrez25}.\\

\appendix
\section{Specific Heat}
\label{appendix}

In the framework of multiscale thermodynamics, a distinction is drawn between microscopic entropy ($dS_0$) and macroscopic entropy ($dS$), rather than between regular and singular entropy.
In the present framework, this decomposition is essential: the thermodynamic fractal dimension $D$ is not intended to describe the full entropy, but rather its scale-dependent singular sector. Therefore, all multiscale contributions associated with Equation~\ref{eq:ms1} should be interpreted as effective representations of $S_{sing}$ under coarse-graining across scales.

Near a continuous phase transition, it is standard within renormalization-group theory that the free energy density admits a decomposition into analytic and non-analytic parts. Consequently, the entropy inherits the structure
\begin{equation}
S=S_n+S_{sing},
\end{equation}
where $S_n$ is analytic in temperature and $S_{sing}$ encodes all non-analytic scaling contributions associated with criticality. Correspondingly, the singular behavior of the thermodynamic potentials near the critical point originates from the non-analytic contribution to the free energy, from which the singular entropy and specific heat are derived. In the standard description of the $2d$ Ising model, the logarithmic divergence of the specific heat originates from the singular entropy contribution,
\begin{equation}
C_H=T\frac{\partial S_{sing}}{\partial T}\propto \ln{|T-T_c|}.
\end{equation}
Therefore, the relevant question is whether Equation~\ref{eq:ms2} can effectively reproduce or approximate the scaling structure associated with $S_{sing}$ near the critical point. Within the present framework, the multiscale entropy variation $dS$ should therefore be interpreted as an effective coarse-grained representation of the singular sector $dS_{sing}$, while $dS_n$ contributes only regular (non-critical) background terms. This implies that $dS$ represents a general quantity, suggesting a deep connection between Equation~\ref{eq:ms2} and the Ising model. Within this interpretation, the singular thermodynamic behavior does not originate from an explicit non-analytic correction added to the entropy itself, but from the divergence of the effective correlation scale $r(T)$ near the critical point. Since the multiscale entropy transfer depends explicitly on the scale factor through $r^{-D}$, the emergence of critical singularities is interpreted as a geometric consequence of the divergence of correlated domains. Then, the challenge lies in understanding how the temperature dependence of $r$, i.e., $r = r(T)$, can give rise to the singular behavior observed in the specific heat.

Near the phase transition, the specific heat $C_H$ is defined as:
\begin{equation}
\label{eq:A1}
    C_H = T\frac{\partial S}{\partial T} \propto \ln{|T - T_c|}.
\end{equation}
This expression exhibits a logarithmic divergence. Now consider modifying Equation~\ref{eq:A1} by introducing an additive constant and a scaling coefficient $\xi$, leading to the generalized expression:
\begin{equation}
\label{eq:A2}
    C_H \propto 1 + \xi\ln{|T - T_c|}.
\end{equation}
Equations~\ref{eq:A1} and \ref{eq:A2}  provides a local effective representation of the logarithmic behavior in the limit of sufficiently small $\xi$. Defining $x = |T - T_c|$ and noting that $x^\xi = e^{\xi \ln(x)}$, we can expand this as a Taylor series for small $\xi$:
\begin{equation}
\label{eq:A3}
    x^\xi \approx 1 + \xi \ln(x).
\end{equation}
This motivates the effective local representation:
\begin{equation}
\label{eq:A4}
    C_H \propto |T - T_c|^\xi.
\end{equation}

Importantly, Equation~\ref{eq:A4} should be interpreted only as a local effective representation of the logarithmic singularity for sufficiently small $\xi$, and not as an exact asymptotic replacement. The logarithmic divergence characteristic of the $2d$ Ising universality class remains fundamentally distinct from a true power-law divergence. Therefore, the present framework does not aim to reproduce the exact logarithmic critical singularity of the $2d$ Ising model from first principles. Instead, it provides an effective scaling representation that preserves the dominant multiscale structure associated with entropy transfer near criticality.

Then, near the phase transition ($T \sim T_c$), the dominant behavior of the entropy derivative becomes
\begin{equation}
\frac{\partial S_{sing}}{\partial T}
\sim
\frac{|T-T_c|^\xi}{T}.
\end{equation}
Since near criticality $T\approx T_c$ may be treated as approximately constant for scaling purposes, this simplifies to

\begin{equation}
\label{eq:A5}
    \frac{\partial S_{sing}}{\partial T} \sim |T - T_c|^\xi.
\end{equation}

On the other hand, from the definition of $k_V$ in Equation~\ref{eq:ms1}, and taking $r_0 = 1$, the singular entropy contribution is phenomenologically modeled as:
\begin{equation}
\label{eq:A6}
    dS_{sing} = T A_0 r^{-D},
\end{equation}
where $A_0 = C_0 / \omega_0$. It is crucial to note that in Equation \ref{eq:A6}, the dominant scaling contribution associated with $S_{sing}$ is effectively encoded in the term $r^{-D}$. This implies that the multiscale framework provides a holistic description of entropy, where the 'normal' and 'singular' regimes are simply different limits of the same multiscale dissipation process. Taking the derivative with respect to $T$ gives:
\begin{equation}
\label{eq:A7}
    \frac{\partial S_{sing}}{\partial T} = A_0 r^{-D} \left[1 - \frac{D T}{r} \frac{dr}{dT} \right].
\end{equation}

Assuming that the relevant scale of the system is governed by the correlation length near criticality, we set
\begin{equation}
r(T) \sim |T - T_c|^\mu,
\end{equation}
where $\mu$ is a scaling exponent. Since the correlation length diverges as $\rho \sim |T-T_c|^{-\nu}$ near criticality, we assume that the effective coarse-grained scale entering Equation~\ref{eq:A7} is proportional to it, i.e.,
\begin{equation}
r(T) \sim \rho(T),
\end{equation}
which motivates the phenomenological identification of scaling exponents $\mu = -\nu$ under the assumption that the dominant entropy-transfer scale is controlled by the correlation length near criticality. For scaling purposes, the sign of $(T-T_c)$ does not affect the critical exponent, and therefore we use
\begin{equation}
\left|\frac{dr}{dT}\right|
\sim
|T-T_c|^{\mu-1}.
\end{equation} Substituting into Equation~\ref{eq:A7} yields:
\begin{equation}
\label{eq:A8}
    \frac{\partial S_{sing}}{\partial T} \sim |T - T_c|^{-\mu D} \left[1 - D T \frac{\mu}{|T - T_c|} \right].
\end{equation}
As $|T-T_c|\rightarrow 0$, the second term inside the brackets diverges as $|T-T_c|^{-1}$, whereas the first term remains finite. Therefore, the second contribution dominates the asymptotic scaling behavior near criticality. Then, near $T \approx T_c$, we have:
\begin{equation}
\label{eq:A9}
    \frac{\partial S_{sing}}{\partial T} \sim -D T_c \mu |T - T_c|^{-\mu D - 1}.
\end{equation}

Comparing this with the scaling form of Equation~\ref{eq:A5} yields the relation:
\begin{equation}
\label{eq:A10}
    \mu = -\frac{\xi + 1}{D}.
\end{equation}

Assuming that the correlation length determines the effective system size, we identify $\mu = -\nu$, which implies:
\begin{equation}
\label{eq:A11}
    \nu = \frac{\xi + 1}{D}.
\end{equation}

In the limit $\xi \to 0$, Equation~\ref{eq:A11} reduces to
\begin{equation}
\nu \sim \frac{1}{D}.
\end{equation}
Using the phenomenological relation proposed in Equation~\ref{eq:5}, namely $d_R \sim 1/\nu$, this leads to the effective correspondence
\begin{equation}
D \sim d_R,
\end{equation}
providing phenomenological consistency with the assumptions introduced before Equation~\ref{eq:cl} and Equation~\ref{eq:ms4}.

By returning to the Equation \ref{eq:A11}:
\begin{equation}
\label{eq:A12}
    \xi = \nu d_R - 1.
\end{equation}
This relation should not be interpreted as an exact exponent identity for the Ising universality class, but rather as an effective scaling constraint emerging from the multiscale thermodynamic framework. This equation also shows that the scaling exponent $\xi$ varies as a function of both $\nu$ and $d_R$, reinforcing the interpretation of Equation~\ref{eq:ms2} as a general framework for describing critical behavior during phase transitions.

{\bf Acknowledgments -} This work was supported by
the Conselho Nacional de Desenvolvimento Cient\'{i}fico e Tecnol\'{o}gico (CNPq), Grant No.  CNPq-01300.008811/2022-51.

\bibliography{referencesV1}

\end{document}